\documentclass[aps,preprint,epsfig,rotate]{revtex4}
\usepackage{graphicx}
\usepackage{bm}
\usepackage{epsfig}

\input epsf
\begin{document}
\title{\bf On highly accurate calculations of the excited $n^1S(L =
           0)-$states in helium atoms}
 \author{Alexei M. Frolov}
 \email[E--mail address: ]{afrolov@uwo.ca}

 \author{David M. Wardlaw}
 \email[E--mail address: ]{dwardlaw@uwo.ca}

\affiliation{Department of Chemistry\\
 University of Western Ontario, London, Ontario N6H 5B7, Canada}

\date{\today}

\begin{abstract}

The total energies and various bound state properties of the excited $2^1S(L
= 0)-$states in two-electron helium atoms, including the ${}^{\infty}$He,
${}^4$He and ${}^3$He atoms, are determined to very high numerical accuracy.
The convergence of the results obtained for some electron-nuclear and
electron-electron expectation values and, in particular, for the
electron-nuclear and electron-electron cusp values, is discussed. The field
component of the isotope shift and lowest order QED correction are estimated
for the $2^1S(L = 0)-$states in the ${}^4$He and ${}^3$He atoms. We also
apply our highly accurate methods to numerical computations of the excited
$n^1S-$states (for $n$ = 3 and 4) in two-electron atomic systems.

PACS number(s): 31.15.ac, 31.15.ae and 31.30.Gs
\end{abstract}
\maketitle

\section{Introduction}\label{intro}

In this communication we consider the bound state properties of the excited
$2^1S(L = 0)-$states in two-electron helium atoms: ${}^{\infty}$He, ${}^4$He
and ${}^3$He. In our previous work \cite{JCP1} we have performed highly
accurate computations for the ground $1^1S(L = 0)-$state and for the triplet
$2^3S(L = 0)-$states in a number of helium-like ions. Another work
\cite{JPB03} contains highly accurate results for the singlet $2^1P(L = 1)-$
and triplet $2^3P(L = 1)-$states in helium atom(s). In this study we want to
consider the excited $2^1S(L = 0)-$states in two-electron helium atoms. Our
current interest to the excited $2^1S(L = 0)-$states in the He atoms stems
from the following. First, many bound state properties of the excited
$2^1S(L = 0)-$states in helium atoms have not been computed in earlier
studies (see, e.g., \cite{Acad}, \cite{Drak} and references therein).

Second, there is a common believe that construction of highly accurate wave
functions for the excited states in two-electron atoms and ions is a
significantly more complicated process than in the case of the lowest states
with the same $L$ and $S$ quantum numbers. Here and below the notations $L$
and $S$ designate the quantum numbers of total angular momentum and electron
spin, respectively (see, e.g., \cite{BS} and \cite{Eps}). Most of the
expected complications most likely arise from various numerical
instabilities which become crucial during optimization of the non-linear
parameters in the wave functions. Indeed, the repeated optimization and
re-optimization of the non-linear parameters in the wave function is a
potentially unstable process for the excited states. Formally, by optimizing
these parameters we are trying to decrease the second, third, $\ldots$,
$n-th$ eigenvalue, i.e. the second, third, etc, root of the eigenvalue
equation. The orthogonality of the wave functions of the excited states to
the wave function of the ground state is not checked during such an
optimization. This means that after a few steps of the optimization of
non-linear parameters the process of construction of a highly accurate wave
function for an excited state may begin to converge to the ground state wave
function. This `instability' problem explains a relatively modest progress
achieved in highly accurate calculations of the excited states in comparison
to the ground state.

Third, it is commonly assumed that, for some unexplained reasons, the
overall convergence rate for bound state properties of the excited atomic
states will be substantially lower than analogous convergence rate for the
same properties determined for the lowest energy state with the same $L$ and
$S$ numbers. This problem has not been studied carefully in earlier works.
It can be answered by computing a large number of bound state properties for
the excited $2^1S-$state in the He atom(s) with the use of different number
of basis functions.

\section{Method}

Our computational goal in this study is to determine the highly accurate
solutions, i.e. the eigenstates and corresponding wave functions, of the
non-relativistic Schr\"odinger equation $H \Psi({\bf r}_1, {\bf r}_2, {\bf
r}_3) = E \Psi({\bf r}_1, {\bf r}_2, {\bf r}_3)$, where $E < 0$ \cite{LLQ},
for $2^1S(L = 0)-$state(s) in the neutral helium atom. For an arbitrary
two-electron atomic system with the nuclear charge $Q e$ the
non-relativistic Hamiltonian is written in the following form
\begin{eqnarray}
 H = - \frac{\hbar^2}{2 m_e} \nabla_1^2 - \frac{\hbar^2}{2 m_e} \nabla_2^2
 - \frac{\hbar^2}{2 m_e} \Bigl(\frac{m_e}{M}\Bigr) \nabla_3^2
 - \frac{Q e^2}{r_{32}} - \frac{Q e^2}{r_{31}} + \frac{e^2}{r_{21}} \; \; ,
 \label{ham1}
\end{eqnarray}
where $r_{ij} = \mid {\bf r}_i - {\bf r}_j \mid = r_{ji}$ ($i \ne j$ = (1,
2, 3)) and ${\bf r}_i$ is the radius-vector of the $i-$th particle written
in Cartesian coordinates. Also, in this equation $\nabla_i = \Bigl(
\frac{\partial}{\partial x_i}, \frac{\partial}{\partial y_i},
\frac{\partial}{\partial z_i} \Bigr)$ is the gradient operator for the
particle with index $i$ ($i$ = 1, 2, 3). Here and everywhere below in this
study the subscripts 1 and 2 designate the two electrons, while the
subscript 3 stands for the nucleus. In the Hamiltonian, Eq.(\ref{ham1}), $Q
= q_3$ is the nuclear charge, while $M = m_3 \gg 1$ is the nuclear mass.
In atomic units where $\hbar = 1, e = 1, m_e = 1$ the Hamiltomian $H$,
Eq.(\ref{ham1}), takes the form
\begin{eqnarray}
 H = - \frac12 \nabla_1^2 - \frac12 \nabla_2^2 - \frac{1}{2 M} \nabla_3^2
 - \frac{Q}{r_{32}} - \frac{Q}{r_{31}} + \frac{1}{r_{21}} \; \; ,
 \label{ham2}
\end{eqnarray}

In this study the unknown wave functions are approximated with the use of
exponential variational expansion in relative coordinates $r_{32}, r_{31}$
and $r_{21}$. For the excited $2^1S(L = 0)-$states, the exponential
variational expansion for the spatial part of an arbitrary two-electron wave
function takes the form \cite{Fro01}
\begin{eqnarray}
 \Psi_{LM} = \frac{1}{\sqrt{2}} (1 + \hat{P}_{21}) \sum_{i=1}^{N}
 C_i \exp(-\tilde{\alpha}_{i} u_1 - \tilde{\beta}_{i} u_2 -
 \tilde{\gamma}_{i} u_3) \label{exp} \\
 = \frac{1}{\sqrt{2}} (1 + \hat{P}_{21}) \sum_{i=1}^{N} C_i
 \exp(-\alpha_{i} r_{32} - \beta_{i} r_{31} - \gamma_{i} r_{21}) \nonumber
\end{eqnarray}
where $C_i$ are the linear (or variational) parameters, $\alpha_i, \beta_i$
and $\gamma_i$ are the non-linear parameters $(i = 1, 2, \ldots, N)$ of
variational expansion, Eq.(\ref{exp}). The operator $\hat{P}_{21}$ is the
permutation of the two identical (1 and 2) particles (electrons) in the
symmetric two-electron ions. In this equation $u_1, u_2, u_3$ are the three
perimetric coordinates which are truly independent and simply related to the
three relative coordinates: $u_i = \frac12 (r_{ik} + r_{ij} - r_{jk})$. The
inverse relation takes the form $r_{ij} = u_i + u_j$, where $(i, j, k) = (1,
2, 3)$. Note that each of the perimetric coordinates varies between 0 and
$+\infty$.

In general, by using the variational expansion, Eq.(\ref{exp}), it is
possible to construct extremely accurate wave functions for the $S(L =
0)-$states in arbitrary three-body systems, if (and only if) the non-linear
parameters $\alpha_i, \beta_i$ and $\gamma_i$ in Eq.(\ref{exp}) are varied.
In this study we shall use the numerical methods and optimization strategy
developed in our earlier work \cite{Fro01}. The procedure developed in
\cite{Fro01} allows one to produce extremely accurate (or essentially exact)
variational results for arbitrary three-body systems, including the Ps$^-$
and H$^-$ ions \cite{Fro07}. Many of these systems, however, have either one
ground bound $1^{1}S-$state, or a few bound states with different $L$ and
$S$ quantum numbers. For two-electron atoms and ions which have infinite
numbers of excited states we modified our strategy developed in \cite{Fro01}
to provide a better numerical stability during each step of optimization.

\section{Properties}

The results of our variational calculations for the excited $2^1S(L =
0)-$state in the helium atom with infinitely heavy nucleus, i.e. in the
${}^{\infty}$He atom, can be found in Table I which contains the total
energies $E$ and $\langle r_{21} \rangle, \langle \delta_{31} \rangle$ and
$\nu_{31}$ expectation values expressed in atomic units. The notations
$\langle r_{21} \rangle, \langle \delta_{31} \rangle$ and $\nu_{31}$ are
explained below. These expectation values have been determined with the use
of trial wave functions with the different number of basis functions $N$. In
this study to approximate the highly accurate wave functions we have used
the exponential variational expansion in relative coordinates $r_{32},
r_{31}$ and $r_{21}$, Eq.(\ref{exp}).

Table II contains the expectation values of many atomic properties of the
$2^1S-$state (expressed in atomic units or $a.u.$) determined using the most
accurate wave functions obtained in this study. The physical meaning of
almost all of the expectation values presented in Tables I and II is clear,
and we make here only a few following remarks. The expectation values of
interparticle distances and their powers are designated as $\langle r^k_{ij}
\rangle$ for $k = 1, 2, 3, 4; -1, -2$ and $(ij)$ = (31), (21) for the
two-electron He atom. The expectation values of the electron-nuclear,
electron-electron and triple delta-functions are designated as $\langle
\delta({\bf r}_{31}) \rangle = \langle \delta_{31} \rangle, \langle
\delta({\bf r}_{21}) \rangle = \langle \delta_{21} \rangle, \langle
\delta({\bf r}_{321}) \rangle = \langle \delta({\bf r}_{32}) \delta({\bf
r}_{31}) \rangle = \langle \delta_{321} \rangle$. Also, Tables I and II
include the electron-nuclear ($\nu_{31}$) and electron-electron ($\nu_{21}$)
cusp values \cite{Kat1}, \cite{Pack}:
\begin{equation}
 \nu_{ij} = \frac{\langle \delta({\bf r}_{ij}) \frac{\partial}
 {\partial r_{ij}} \rangle}{\langle \delta({\bf r}_{ij}) \rangle}
\end{equation}
where $\delta({\bf r}_{ij}) = \delta_{ij}$ is the appropriate
delta-function and $(ij)$ = (31), (21). In this study we shall assume that
all point particles interact with each other by the Coulomb potentials.
Therefore, the expected (exact) two-particle cusp equals $\nu_{ij} = q_i
q_j \frac{m_i m_j}{m_i + m_j}$ \cite{Kat1}, \cite{Pack}, where $q_i, q_j$
are the particle's charges and $m_i, m_j$ are their masses. For the
considered singlet $2^1S(L = 0)-$states in the two-electron helium atom with
the infinitely heavy nucleus ${}^{\infty}$He, the expected electron-nuclear
cusp equals $-Q = -2$, while the electron-electron cusp equals 0.5.
For the ${}^3$He and ${}^4$He helium atoms the electron-electron cusps also
equal 0.5 (exactly), while electron-nuclear cusps include some small
correction for the $\frac{m_e}{M} = \frac{1}{M}$ ratio. For the ${}^3$He and
${}^4$He helium atoms one finds for the electron-nuclear cusp (in atomic
units)
\begin{equation}
 \nu_{31} = - Q \frac{M}{M+ 1} = -Q \Bigl( 1 - \frac{1}{M + 1} \Bigr)
\end{equation}
where $M \gg 1$ and $Q$ is the nuclear electric charge. In our calculations
of the $2^1S-$states in the ${}^3$He and ${}^4$He helium atoms the following
values of nuclear mass were used: $M({}^{3}$He) = 5495.8852 $m_e$ and
$M({}^{4}$He) = 7294.2996 $m_e$ \cite{COD}, \cite{CRC}.

The coincidence between the expected and computed cusp values provides a
very convenient, accurate and universal criterion to control the overall
quality of the wave function constructed in our computations. In general,
the electron-electron cusp is a better criterion of the overall quality of
the variational wave function than the electron-nuclear cusp. This means
that in actual calculations it is much harder to obtain a good/excellent
agreement for the electron-electron cusp, than a similar agreement for the
electron-nuclear cusp.

Table II also contains the expectation values of the two interparticle
$cosine-$functions which are determined as follows
\begin{eqnarray}
\tau_{ij} = \langle \cos ({\bf r}_{ik} {}^{\wedge}{\bf r}_{jk}) \rangle =
\langle \frac{{\bf r}_{ik} \cdot {\bf r}_{jk}}{r_{ik} r_{jk}} \rangle
\; \; \; ,
\end{eqnarray}
where $(i,j,k) = (1,2,3)$. The $\tau_{ij}$ expectation values are always
$\le 1$. The absolute value of $\tau_{21} (\equiv \tau_{12})$ can be
considered as an `ideal' measure of the static electron-electron
correlations in the two-electron atomic systems. Let us define the quantity
$\langle f \rangle$ which is expressed in terms of the relative coordinates
($r_{31}, r_{32}, r_{21}$) or perimetric coordinates ($u_{1}, u_{2}, u_{3}$)
as follows:
\begin{eqnarray}
 \langle f \rangle = \frac12 \langle \psi \mid \frac{u_{1}}{r_{32}}
 \frac{u_{2}}{r_{31}} \frac{u_{3}}{r_{21}} \mid \psi \rangle =
 \int\!\! \int\!\! \int \mid \psi (u_{1},u_{2},u_{3}) \mid^{2}
 u_{1} u_{2} u_{3} d u_{1} d u_{2} d u_{3} \; \; \; . \label{ff}
\end{eqnarray}
It can be shown that the equality
\begin{equation}
 \tau_{21} + \tau_{32} + \tau_{31} = 1 + 4 \langle f \rangle
\end{equation}
holds for arbitrary three-body system. For the two-electron (i.e. symmetric)
ions/atoms we always have $\tau_{32} = \tau_{31}$, and therefore, $\tau_{21}
+ 2 \tau_{31} = 1 + 4 \langle f \rangle$. It can be also shown that in an
arbitrary Coulomb three-body system $0 \le \langle f \rangle < 0.085$. The
$\langle f \rangle$ value can be calculated either directly from
Eq.(\ref{ff}), or by applying the expectation values of the $cosine$
functions $\tau_{ij}$ computed earlier. The coincidence of these two values
of $\langle f \rangle$ indicates that the $\tau_{32}, \tau_{31}, \tau_{21}$
and $\langle f\rangle$ expectation values have been computed correctly.

The virial factor $\eta$ in Table II is determined as follows:
\begin{equation}
 \eta = \mid 1 + \frac{\langle V \rangle}{2 \langle T \rangle} \mid
\end{equation}
where $\langle T \rangle$ and $\langle V \rangle$ are the expectation values
of the kinetic and potential energy, respectively. The deviation
of the factor $\eta$ from zero indicates, in principle, the overall quality
of the variational wave function used \cite{Fock}. In particular, for the
wave functions used in our present calculations, the virial parameters
$\eta$ is in the range $1 \cdot 10^{-18} - 5 \cdot 10^{-19}$, showing that
our wave functions are highly accurate.

Note that some of the bound state properties from Table II can be expressed
as the linear combinations of other properties. For instance, by using the
identity
\begin{equation}
 {\bf r}_{31} = {\bf r}_{32} - {\bf r}_{21}
\end{equation}
one finds for the $\langle {\bf r}_{31} \cdot {\bf r}_{32} \rangle$ and
$\langle {\bf r}_{31} \cdot {\bf r}_{32} \rangle$ expectation values
\begin{equation}
 \langle {\bf r}_{31} \cdot {\bf r}_{32} \rangle = \frac12 (\langle
 r^2_{32} \rangle + \langle r^2_{31} \rangle - \langle r^2_{21} \rangle)
\end{equation}
and
\begin{equation}
 \langle {\bf r}_{21} \cdot {\bf r}_{32} \rangle = \frac12 (\langle r^2_{32}
 \rangle + \langle r^2_{21} \rangle - \langle r^2_{31} \rangle)
\end{equation}
here and everywhere below the notation ${\bf a} \cdot {\bf b}$ stands for
the scalar product of the ${\bf a}$ and ${\bf b}$ vectors, i.e. ${\bf a}
\cdot {\bf b} = a_x b_x + a_y b_y + a_z b_z$, where ${\bf a} = (a_x, a_y,
a_z)$ and ${\bf b} = (b_x, b_y, b_z)$. Analogously, since $({\bf p}_1 + {\bf
p}_2 + {\bf p}_3) \mid \Psi \rangle = 0$, we have
\begin{equation}
 \langle {\bf p}_{i} \cdot {\bf p}_{j} \rangle = \frac12 (\langle p^2_{i}
 \rangle + \langle p^2_{j} \rangle - \langle p^2_{k} \rangle) \label{e11}
\end{equation}
where $(i,j,k) = (1,2,3)$. In the gradient form this equality takes the form
\begin{equation}
 \langle \nabla_i \cdot \nabla_j \rangle = \frac12 (\langle \nabla^2_i
 \rangle + \langle \nabla^2_j \rangle - \langle \nabla^2_k \rangle) =
 - \langle - \frac12 \nabla^2_i \rangle - \langle - \frac12 \nabla^2_j
 \rangle + \langle - \frac12 \nabla^2_k \rangle \label{e12}
\end{equation}
where $\langle - \frac12 \nabla^2_i \rangle$ ($i$ = 1, 2, 3) are the
single-particle kinetic energies. In the general case, the both sides of the
equalities Eqs.(\ref{e11}) - (\ref{e12}) can be computed separately, i.e.
such relations can be also used to control the overall quality of the
variational wave function. The $\langle {\bf p}_1 \cdot {\bf p}_2 \rangle$
expectation value can also be used as a measure of the dynamical
electron-electron correlation in the two-electron ions. It should be noticed
that there is an obvious difference between the $\langle {\bf p}_1 \Psi \mid
{\bf p}_2 \Psi \rangle$ and $\langle \Psi \mid {\bf p}_1 \cdot {\bf p}_2
\mid \Psi \rangle$ expectation values. In fact, these two expectation values
differ from each other by sign, i.e. $\langle {\bf p}_1 \Psi \mid {\bf p}_2
\Psi \rangle = -\langle \Psi \mid {\bf p}_1 \cdot {\bf p}_2 \mid \Psi
\rangle$.

\subsection{Singular expectation values}

All expectation values mentioned above are regular, i.e. their analytical
and/or numerical computation is relatively simple. A few expectation values
from Table II, however, contain some singular parts. In such cases some
additional explanations are needed. For instance, consider the expectation
value $\langle \frac{1}{r^{3}_{32}} \rangle = \langle \frac{1}{r^{3}_{31}}
\rangle$. In relative coordinates $r_{32}, r_{31}, r_{21}$ the computation
of this expectation value is reduced to the calculation of the following
Hylleraas-type integrals
\begin{eqnarray}
 \Gamma_{-2,1,1}(a,b,c) = \int_0^{\infty} \int_0^{\infty}
 \int_{\mid r_{32}-r_{31} \mid}^{r_{32}+r_{31}} exp(-a r_{32} -b r_{31}
 -c r_{21}) r^{-2}_{32} r_{31} r_{21} dr_{32} dr_{31} dr_{21} \label{int2x}
\end{eqnarray}
for many different sets of $(a,b,c)-$values. Formally, each of these
integrals diverges, i.e. it does not exist as a finite expression. To make
this integral finite one needs to introduce a small (and positive) cutoff
parameter $\epsilon$
\begin{eqnarray}
 \Gamma_{-2,1,1}(a,b,c;\epsilon) = \int_{\epsilon}^{\infty} \int_0^{\infty}
 \int_{\mid r_{32}-r_{31} \mid}^{r_{32}+r_{31}} exp(-a r_{32} -b r_{31}
 -c r_{21}) r^{-2}_{32} r_{31} r_{21} dr_{32} dr_{31} dr_{21} \label{int3}
\end{eqnarray}
Now, this integral is finite for $\epsilon > 0$, but diverges when $\epsilon
\rightarrow 0$. For the finite integral $\Gamma_{-2,1,1}(a,b,c;\epsilon)$
defined by Eq.(\ref{int3}) for $\epsilon > 0$ we can write
\begin{eqnarray}
 \Gamma_{-2,1,1}(a, b, c; \epsilon) = \frac{\partial^2
 \Gamma_{-2,0,0}(a, b, c; \epsilon)}{\partial b \partial c}
\end{eqnarray}
where
\begin{eqnarray}
 \Gamma_{-2,0,0}(a,b,c;\epsilon) = \int_{\epsilon}^{\infty} \int_0^{\infty}
 \int_{\mid r_{32}-r_{31} \mid}^{r_{32}+r_{31}} exp(-a r_{32} -b r_{31}
 -c r_{21}) r^{-2}_{32} dr_{32} dr_{31} dr_{21} \label{int3x}
\end{eqnarray}

The $\Gamma_{-2,0,0}(a, b, c; \epsilon)$ integral is represented as the sum
of its regular ($R$) and singular ($S$) parts, i.e. $\Gamma_{-2,0,0}(a, b,
c; \epsilon) = R_{-2,0,0}(a, b, c) + S_{-2,0,0}(a, b, c; \epsilon)$, where
\begin{eqnarray}
 S_{-2,0,0}(a, b, c; \epsilon) = \frac{2}{b + c} \Bigl( \psi(1) - ln
 \epsilon \Bigr) = - \frac{2}{b + c} (\gamma_E + ln \epsilon) \label{s-2} \\
 R_{-2,0,0}(a, b, c) = 2 \frac{(a + c) ln (a + c) - (a + b) ln (a +
 b)}{(b^2 - c^2)} + \frac{2}{b + c} \label{r-2} \\
 &=& R^{(ln)}_{-2,0,0}(a, b, c) + F_{-2,0,0}(a,b,c) \nonumber
\end{eqnarray}
where $\psi(n)$ is the digamma function \cite{AS} (or $psi-$function defined
in Eq.(8.360) from \cite{GR}). Note that $\psi(1) = -\gamma_E$, where
$\gamma_E \approx 0.577215\ldots$ is the Euler's constant. The
$R^{(ln)}_{-2,0,0}(a, b, c)$ term is the `logarithmic' term which contains
the $ln (a + c)$ and $ln (a + b)$ expressions, i.e.
\begin{eqnarray}
 R^{(ln)}_{-2,0,0}(a,b,c) = \frac{2}{(b^2 - c^2)} \Bigl[(a+c) ln(a+c) -
 (a+b) ln(a+b)\Bigr] \; \; , \nonumber
\end{eqnarray}
while
\begin{eqnarray}
 F_{-2,0,0}(a,b,c) = \frac{2}{b + c} \nonumber
\end{eqnarray}

The second order derivative of the regular part is
\begin{eqnarray}
 R_{-2,1,1}(a, b, c) = \frac{\partial^2 R^{(ln)}_{-2,0,0}(a, b, c)}{\partial
 b \partial c} + \frac{4}{(b + c)^3}
\end{eqnarray}
The expectation values of both sides of this equation are
\begin{eqnarray}
 \langle \Psi \mid R_{-2,1,1}(a, b, c) \mid \Psi \rangle =
 \langle \Psi \mid \frac{\partial^2 R^{(ln)}_{-2,0,0}(a, b, c)}{\partial b
 \partial c} \mid \Psi \rangle + \langle \Psi \mid \frac{4}{(b + c)^3}
 \mid \Psi \rangle
\end{eqnarray}
In the exponential basis the $\frac{4}{(b + c)^3}$ matrix elements
correspond to the $\delta({\bf r}_{32})$ delta-function multiplied by a
factor of $4 \pi$. The corresponding expectation value is
\begin{eqnarray}
 \langle \Psi \mid R_{-2,1,1}(a, b, c) \mid \Psi \rangle =
 \langle \Psi \mid \frac{\partial^2 R^{(ln)}_{-2,0,0}(a, b, c)}{\partial b
 \partial c} \mid \Psi \rangle + 4 \pi \langle \delta({\bf r}_{32}) \rangle
 \label{e23}
\end{eqnarray}
Analogously, the expectation value of the singular part is
\begin{equation}
 \langle \Psi \mid S_{-2,1,1}(a, b, c; \epsilon) \mid \Psi \rangle =
 - 4 \pi \langle \delta({\bf r}_{32}) \rangle (\gamma_E + ln \epsilon)
 \label{s-4}
\end{equation}
where $\langle \delta({\bf r}_{32}) \rangle$ is the expectation value of the
$\delta({\bf r}_{32})$ delta-function.

As follows from the definition $\Gamma_{-2,1,1}(a, b, c; \epsilon) =
R_{-2,1,1}(a, b, c) + S_{-2,1,1}(a, b, c; \epsilon)$ the sum
\begin{equation}
 \Gamma_{-2,1,1}(a, b, c; \epsilon) + S_{-2,1,1}(a, b, c; \epsilon) =
 R_{-2,1,1}(a, b, c) \label{e25}
\end{equation}
is the regular expression and it has the finite limit when $\epsilon
\rightarrow 0$. The limits of both sides of Eq.(\ref{e25}) are
\begin{eqnarray}
 \langle \frac{1}{r^3_{32}} \rangle = \lim_{\epsilon
 \rightarrow 0} \Bigl[ \langle \Psi \mid \frac{1}{r^3_{32}} \mid \Psi
 \rangle_{\epsilon} + 4 \pi \langle \delta({\bf r}_{32}) \rangle (\gamma_E
 + ln \epsilon) \Bigr] = 4 \pi \langle \delta({\bf r}_{32}) \rangle +
 \langle \Psi \mid \frac{1}{r^3_{32}} \mid \Psi \rangle_R
\end{eqnarray}
This expression is considered as the expectation value of the
$\frac{1}{r^3_{32}}$ operator, i.e. we have
\begin{equation}
 \langle \frac{1}{r^3_{32}} \rangle = 4 \pi \langle \delta({\bf r}_{32})
 \rangle + \langle \Psi \mid \frac{1}{r^3_{32}} \mid \Psi \rangle_R
\end{equation}
Note that in contrast with the expectation values of regular operators the
$\langle \frac{1}{r^3_{32}} \rangle$ does not coincide with the $\langle
\Psi \mid \frac{1}{r^3_{32}} \mid \Psi \rangle_R$ expectation value. The
presence of the finite difference $4 \pi \langle \delta({\bf r}_{32})
\rangle$ is typical for the expectation values of singular operators. For
the $\langle \frac{1}{r^3_{21}} \rangle$ expectation value the analogous
formula takes the form \cite{Fro5}
\begin{eqnarray}
 \langle \frac{1}{r^3_{21}} \rangle = \lim_{\epsilon
 \rightarrow 0} \Bigl[ \langle \Psi \mid \frac{1}{r^3_{21}} \mid \Psi
 \rangle_{\epsilon} + 4 \pi \langle \delta({\bf r}_{21}) \rangle (\gamma_E
 + ln \epsilon) \Bigr] = 4 \pi \langle \delta({\bf r}_{21}) \rangle +
 \langle \Psi \mid \frac{1}{r^3_{21}} \mid \Psi \rangle_R
\end{eqnarray}

Note also that in Eq.(\ref{e23}) the second order derivative is
\begin{eqnarray}
 \frac{\partial^2 R^{(ln)}_{-2,0,0}(a, b, c)}{\partial b \partial c} =
 16 \frac{[(a+c) ln(a + c) - (a + b) ln(a + b)] b c}{(c^2 - b^2)^3}
 \label{e21} \\
 - 4 \frac{[ln(a + b) + 1] c}{(b^2 - c^2)^2} - 4 \frac{[ln(a + c) + 1]
 b}{(b^2 - c^2)^2} \nonumber
\end{eqnarray}
It is easy to understand that this formula is computationally unstable,
when $c \rightarrow b$. Indeed, both the numerator and denominator in this
formula $\rightarrow 0$ when $c \rightarrow b$. This produces some troubles
in actual calculations. The transformation of this formula to the
computationally stable form when $c \rightarrow b$ can be found in
\cite{Fro5}. The most detailed analysis of various singular integrals
arising in two-electron atomic problems can be found in \cite{HFS}.

\section{Corrections to the total energy}

The expectation values from Table II can be used to determine some actual
properties of the helium atoms in the $2^1S-$state. Here by the `actual
properties' we mean some linear combinations of our expectation values
which can be measured in modern experiments. The most important of such
properties are various lowest order corrections to the total
non-relativistic energies. Formally, by computing all possible lowest order
corrections, e.g., relativistic and quantum electrodynamics corrections,
mass corrections, etc, we must obtain the exact agreement with the energies
measured in high precision experiments. In this Section we consider a few
such corrections to the non-relativistic atomic energies of the
$2^1S-$state in the two-electron helium atom(s).

First, let us evaluate the field component of the total isotope shift for
the $2^1S-$state in helium atoms. The field shift is related to the extended
nuclear charge distribution which produces the non-Coulomb field at
distances close to the nucleus. It is clear that the largest deviations
between the Coulomb and actual potentials can be found close to the atomic
nucleus, i.e. for distances $r \approx r_e \ll \Lambda \ll a_0$, where $r_e
= \alpha^2 a_0$ is the classical electron radius and $\Lambda = \alpha a_0$
is the Compton wave length. Here and below, $\alpha = 7.297352568 \cdot
10^{-3}$ is the fine structure constant and $a_0 \approx 5.29177249 \cdot
10^{-11}$ $m$ is the Bohr radius. Note that all numerical values for the
physical constants used in this study were chosen from \cite{COD},
\cite{CRC}. The general theory of the field shift has been discussed
extensively in a number of works (see, e.g., \cite{Sob}, \cite{FrFi},
\cite{King} and references therein).

In our earlier work \cite{Fro06} we obtained the following expression for
the field shift (in atomic units) of the bound $S(L = 0)-$states in light
atoms and ions (with $Q \le 6$)
\begin{eqnarray}
 E^{fs}_M = \frac{8 \pi}{3} Q \rho_e(0) R^2 \Bigl(\frac{3 + \lambda}{5 -
 \lambda} \Bigr) = \frac{8 \pi}{5} Q \rho_e(0) R^2 \Bigl( \frac{1 + \frac13
 \lambda}{1 - \frac15 \lambda} \Bigr) = \frac{8 \pi}{5} Q \alpha^4 \cdot
 \langle \delta({\bf r}_{e N}) \rangle \Bigl(\frac{R}{r_e}\Bigr)^2 \xi
 \label{e23A}
\end{eqnarray}
where $Q$ is the nuclear charge and $R$ is the nuclear radius. Our formula
for the field shift $E^{fs}_M$ follows from the well known expression
obtained by Racah, Rosental and Breit (see, e.g., \cite{Sob}). Its explicit
derivation can be found in the Appendix. The parameter $\lambda$ and related
factor $\xi = \frac{1 + \frac13 \lambda}{1 - \frac15 \lambda}$ in
Eq.(\ref{e23A}) describe the actual charge/proton distribution in the
nucleus. Also, in this expression $r_e = \frac{e^2}{m_e c^2} = \alpha^2 a_0
\approx$ 2.81794093 $fm$ (1 $fm$ (fermi) = $1 \cdot 10^{-13}$ $cm$) is the
classical electron radius. In general, the nuclear radius $R \approx r_e$
and its actual value depends upon the total number of nucleons $A$ in the
nucleus ($R \sim A^{\frac13}$). In other words, the field shift formally
corresponds to the $\alpha^4-$correction to the energy levels, i.e. to the
second order relativistic correction. Let us evaluate the $E^{fs}_M$ shifts
for the $2^1S-$state in the ${}^3$He and ${}^4$He atoms. The nuclear sizes
used in our computations were $R({}^{3}$He) = 1.880 $fm$ \cite{McCar} and
$R({}^{4}$He) = 1.6773 $fm$ \cite{Car} (see also \cite{Elt} and \cite{Bar}).
We have also selected zero value for the parameter $\lambda$, i.e. $\xi$ = 1
in Eq.(\ref{e23A}). This means that the uniform (or $r-$independent) proton
density distribution over the volume of the nucleus is assumed for each of
the ${}^{3}$He and ${}^{4}$He nuclei. This assumption produces the following
values for the field shifts $E^{fs}_M({}^{3}$He) = 1.66061$\cdot 10^{-8}$
$a.u.$ and $E^{fs}_M({}^{4}$He) = 1.32120$\cdot 10^{-8}$ $a.u$. Numerical
recalculation of these values to $cm^{-1}, MHz, eV$ and other units is
straightforward.

Another example is the lowest order ($\sim \alpha^3$) QED correction to the
non-relativistic energies of the $2^1S-$state of the helium atoms. For an
arbitrary bound $S(L = 0)-$state in the ${}^{\infty}$He atom the closed
analytical formula for the lowest order QED correction is written in the
form (in atomic units)
\begin{eqnarray}
 \Delta E^{(3)} = \frac{8}{3} Q \alpha^3 \Bigl[ \frac{19}{30} -
 2 \ln \alpha - \ln K_0 \Bigr] \langle \delta({\bf r}_{31}) \rangle +
 \alpha^3 \Bigl[ \frac{164}{15} + \frac{14}{3} \ln \alpha
 \label{eq15} \\
 - \frac{10}{3} S(S + 1) \Bigr] \langle \delta({\bf r}_{21}) \rangle
 - \frac{14}{3} \alpha^3 \Bigl(\frac{1}{4 \pi} \langle \frac{1}{r^3_{21}}
 \rangle \Bigr) \nonumber
\end{eqnarray}
where $\alpha$ is the fine structure constant, $Q$ (= 2) is the nuclear
charge and $S$ is the total spin of two electrons. For the singlet states we
always have $S = 0$. Also, in this formula $\ln K_0$ is the Bethe logarithm
\cite{BS}, which is represented in the form $\ln K_0 = \ln k_0 + 2 \ln Q$,
where $\ln k_0$ is the charge-reduced Bethe logarithm. Our current numerical
evaluation of the charge-reduced Bethe logarithm for the $2^1S-$state of the
${}^{\infty}$He atom is $\ln k_0 \approx 2.98011831$. This value is not very
accurate and must be improved in future calculations. Therefore, the value
of Bethe logarithm for this state is $\ln K_0 \approx$ 4.36641267. By using
this value of $\ln K_0$ and expectation values from Table II for the one
finds that for the ${}^{\infty}$He atom $\Delta E^{(3)} \approx 1.1890905
\cdot 10^{-5}$ $a.u.$ This is the lowest order ($\sim \alpha^3$) QED
correction determined for the $2^1S-$state in the ${}^{\infty}$He atom.

For the two-electron helium atoms with the finite nuclear masses we need to
determine the corresponding finite mass correction (or recoil correction,
for short). The recoil correction to the lowest order QED correction in the
case of $S(L = 0)-$states in two-electron atoms/ions is represented in the
form \cite{PuSa} (in atomic units):
\begin{eqnarray}
 \Delta E^{(3)}_M = \Bigl(\frac{M}{M + 1} - \frac{2}{M}\Bigr) \Delta
 E^{(3)} + \frac{4 \alpha^3}{3 M} \Bigl[ \frac{31}{3} + 2 - \ln \alpha - 4
 \ln K_0 \Bigr] \langle \delta({\bf r}_{31}) \rangle \label{eq16} \\
 -  \frac{2}{M} \cdot \frac{14}{3} \alpha^3 \Bigl(\frac{1}{4 \pi}
 \langle \frac{1}{r^3_{31}} \rangle \Bigr) \nonumber
\end{eqnarray}
where $\Delta E^{(3)}$ is the expression from Eq.(\ref{eq15}), while $M$ ($M
\gg 1$) is the nuclear mass (expressed in the electron mass $m_e$). By using
our expectation values from Table II and charge-reduced Bethe logarithm
determined for the $2^1S-$state in the ${}^{\infty}$He atom one finds from
Eq.(\ref{eq16}) that $\Delta E^{(3)}_M({}^{4}$He$) \approx 1.1891233 \cdot
10^{-5}$ $a.u.$ and $\Delta E^{(3)}_M({}^{3}$He$) \approx 1.1891340 \cdot
10^{-5}$ $a.u.$ To obtain a slightly better accuracy in the last equation we
can use the expectation values from Table II determined for the two-electron
He atoms with the finite nuclear masses. In these cases one finds from
Eq.(\ref{eq16}) $\Delta E^{(3)}_M({}^{4}$He$) \approx 1.1891230 \cdot
10^{-5}$ $a.u.$ and $\Delta E^{(3)}_M({}^{3}$He$) \approx 1.1891319 \cdot
10^{-5}$ $a.u.$

The last example which we want to consider here is related to the Vinti
identity \cite{Vint}. For two-electron atoms/ions this identity takes the
form
\begin{eqnarray}
 \langle {\bf p}_1 \cdot {\bf p}_2 \rangle = \frac{Q}{2} \langle {\bf
 r}_{31} \cdot {\bf r}_{32} (\frac{1}{r^3_{31}} + \frac{1}{r^3_{32}})
 \rangle + \frac12 \langle \frac{1}{r_{12}} \rangle = Q \langle
 \frac{{\bf r}_{31} \cdot {\bf r}_{32}}{r^3_{31}} \rangle + \frac12 \langle
 \frac{1}{r_{12}} \rangle
\end{eqnarray}
The $\langle \frac{{\bf r}_{31} \cdot {\bf r}_{32}}{r^3_{31}} \rangle =
\frac12 \Bigl[ \langle \frac{1}{r_{31}} \rangle + \Bigl( \langle
\frac{r^2_{32}}{r_{31}} \rangle - \langle \frac{r^2_{32}}{r_{31}} \rangle
\Bigr)\Bigr]$ expectation value can be computed either directly (see Table
II), or with the use of the following relation
\begin{eqnarray}
 \langle \frac{{\bf r}_{32} \cdot {\bf r}_{31}}{r^3_{32}} \rangle =
 \frac{1}{Q} \langle p^2_1 \rangle - \frac{1}{2 Q} \langle p^2_3 \rangle
 - \frac{1}{2 Q} \langle \frac{1}{r_{12}} \rangle \label{last}
\end{eqnarray}
since $\langle {\bf p}_1 \cdot {\bf p}_2 \rangle = \frac12 \langle p^2_{1}
\rangle + \frac12 \langle p^2_{2} \rangle - \frac12 \langle p^2_{3} \rangle$
and $\langle p^2_1 \rangle = \langle p^2_2 \rangle$ in any two-electron
system. All expectation values mentioned in the right-hand side of this
equality can also be found in Table II. Analogous relations can be derived
for the $\langle \frac{{\bf r}_{31} \cdot {\bf r}_{21}}{r^3_{31}} \rangle$
expectation value. Note that all expectation values in the right-hand side
of Eq.(\ref{last}) are regular, while the expectation value in the left-hand
side of Eq.(\ref{last}) contains the difference of the two singular
expectation values $\langle \frac{r^2_{31}}{r^3_{32}} \rangle$ and $\langle
\frac{r^2_{21}}{r^3_{32}} \rangle$. As follows from Eq.(\ref{last}) such a
difference of these two singular expectation values is a regular value. This
means that the two singular parts cancel each other completely. In general,
an accurate numerical coincidence of the $\langle \frac{{\bf r}_{31} \cdot
{\bf r}_{32}}{r^3_{31}} \rangle$ expectation values computed directly and
with the use of Eq.(\ref{last}) is another important test for our highly
accurate wave functions. An analogous test can be used for the $\langle
\frac{{\bf r}_{31} \cdot {\bf r}_{21}}{r^3_{31}} \rangle$ expectation value.
It is also interesting to note that for the singlet states in two-electron
ions the $\langle \frac{{\bf r}_{31} \cdot {\bf r}_{32}}{r^3_{31}} \rangle$
expectation values slowly vary with the nuclear charge $Q$. Moreover, these
expectations values have a finite limit ($\approx -0.1755$) when $Q
\rightarrow \infty$.

\section{Conclusion}

We have performed highly accurate computations of the excited $2^1S(L =
0)-$states in the two-electron helium atoms: ${}^{\infty}$He, ${}^4$He and
${}^3$He. The total energies and a large number of bound state properties
have been determined for the $2^1S-$state in the ${}^{\infty}$He, ${}^4$He
and ${}^3$He atoms to very high numerical accuracy. By using our highly
accurate variational wave functions we have also evaluated the expectation
values of some singular operators. To the best of our knowledge this work
is the first extensive study of the bound state properties of the excited
$2^1S(L = 0)-$states in two-electron helium atoms and most of the bound
state properties computed in this work have never been evaluated in earlier
studies. The knowledge of accurate expectation values given In Table II
allows us to determine various lowest order relativistic, QED and mass
corrections to the total energies of the $2^1S-$state(s) in the
${}^{\infty}$He, ${}^4$He and ${}^3$He atoms.

In our extensive numerical calculations of the excited $n^1S-$states of the
He atom(s) we have found no additional complications for our highly accurate
procedure which is based on careful optimization of many non-linear
parameters at small and intermediate dimensions. Briefly, this means that
`instability problem' mentioned in the Introduction is not critically
important for our method. As follows from computational results obtained in
this study our methods can also be used for highly accurate computations of
the excited states in two-electron atoms and ions. In fact, we have
performed some of such calculations. Table III contains our preliminary
results for the total energies of the $3^1S-$ and $4^1S-$states in the
${}^{\infty}$He atom. As follows from Table III the total energies of the
$3^1S$ and $4^1S$ bound states converge very fast. Finally, our results from
Table III for the the $3^1S-$ and $4^1S-$states in the ${}^{\infty}$He atom
are the most accurate total energies ever obtained for these states. It
shows a great potential of our method for highly accurate computations of
bound states in two-electron systems. In our test calculations of the bound
$5^1S-$state in the ${}^{\infty}$He atom the short term booster function has
not been constructed. Optimization of the three boxes for the non-linear
parameters \cite{Fro01} allows us to produce a very compact (100-term) wave
function which corresponds to the energy -2.0211768515145 $a.u.$ The
difference with the `exact' energy for this state is $\approx 6.5 \cdot
10^{-11}$ $a.u.$ Very likely, that further optimization of the non-linear
parameters in the wave functions with 200 basis functions, Eq.(\ref{exp}),
will produce the variational energy which is very close to the known energy
for this state. Analogous situation can be found for other highly excited
$S-, P-$ and $D-$states in the ${}^{\infty}$He atom. In future studies we
want to develop the new optimization strategy which can be used to chose the
non-linear parameters in trial wave functions for highly excited $S-, P-$
and $D-$ states. This will allow us to produce very compact and highly
accurate wave functions for these states. Computations with the use of very
large wave functions ($\ge$ 3000 basis functions) can be avoided for highly
excited bound states in two-electron atoms/ions. However, the current
situation with highly accurate computations of highly excited singlet
$S-$states (e.g., for the $8^1S-, 9^1S-, 10^1S-$ and higher singlet
$S-$states) in the helium atoms and helium-like ions is far from
satisfaction for our method, since the Hylleraas variational expansion (see,
e.g., \cite{Drak}) still provides better accuracy for such states (if
comparable numbers of basis functions are used in both methods).

Note that our methods allow one to determine the total energies and many
other bound state properties to very high numerical accuracy which is quite
comparable and even better than analogous accuracy achieved for the ground
states. The overall convergence rates observed in calculations of many bound
state properties of the excited $2^1S(L = 0)-$states in two-electron helium
atoms are relatively high. Briefly, this means that our optimization
strategy used to optimize the non-linear parameters in highly accurate wave
functions works very well for both ground and excited states in two-electron
atoms/ions. This also contains the answer to the third problem mentioned in
the Introduction that is the convergence rate for most of the bound state
properties of the excited states is not substantially lower than for
analogous properties of the ground state. Furthermore, the same optimization
procedure has been applied to highly accurate computations of the $2^{1}P(L
= 1)-$ and $2^{3}P(L = 1)-$states in the He atom(s). The explicit form of
the trial variational wave function for the states with $L \ge 1$ is given
in \cite{Fro01}. The best variational (total) energies obtained in our
computations for these states in the ${}^{\infty}$He atom are:
-2.1238430864981013590742 $a.u.$ and -2.13316419077928320510251 $a.u.$,
respectively \cite{FrWa2010}. Numerical uncertainties in these values can be
evaluated as $\approx 3.5 \cdot 10^{-20}$ $a.u.$ In other words, these
energies and corresponding wave functions are significantly more accurate
than known from the modern literature (see, e.g., \cite{JPB03}).

Our highly accurate method can also be applied to highly accurate
calculations of the weakly-bound singlet and triplet excited states in
two-electron atoms and ions. These are the $n^{1,3}S-, n^{1,3}P-,
n^{1,3}D-$states with $n \ge 2$. Results of these and other similar
computations indicate clearly that clusterization of the wave functions
plays a very important role for the excited states in two-electron atomic
systems. The effect of clusterization for the exponential variational
expansion was originally discovered in \cite{OpSp}. For our two-stage method
\cite{Fro01} the clusterization of variational expansion, Eq.(\ref{exp}),
means the presence of two following things. First, the short term (booster)
wave function produces almost `exact' energy already for $N_0$ = 200 - 300.
For instance, for the $3^1S(L = 0)-$state the booster function with only
$N_0$ = 200 exponential basis functions allows one to obtain the total
energy $E$ = -2.061271989738157 $a.u.$ (current value), while the `exact'
energy of this state in the ${}^{\infty}$He atom is -2.061271989740911
$a.u$. Second, optimization performed at the second stage of our method
produces a number (usually three) of `optimal' parallelotops which are used
later to chose parameters $\alpha_i, \beta_i$ and $\gamma_i$ (where $N_0 < i
\leq N$) in Eq.(\ref{exp}). In actual calculations of the excited
$n^{1,3}S-$ and $n^{1,3}P-$states in the He atom(s) for $n \geq 3$, such an
optimization of three parallelotops can be performed for $N$ = 600 in the
total wave function. It produces almost the exact value of the bound state
energy. Moreover, the optimal parallelotops for these bound states have
almost degenerated structure. Briefly, this means that such parallelotops
generate the wave function, Eq.(\ref{exp}), in which all parameters
$\gamma_i$ either equal zero exactly, or very close to zero. But this means
that the role of electron-electron correlations (or $r_{21}$ coordinate)
rapidly decreases for highly excited bound states in two-electron ions
\cite{OpSp}. In respect with this, the difference between the total
energies of the triplet and singlet states with large angular moment $L$ in
He-like atoms and ions rapidly (almost exponentially) decreases, if the
value of $L$ increases. This remarkable property of the exponential
variational expansion, Eq.(\ref{exp}), in applications to Rydberg states has
been observed in all helium-like atoms and ions. It can be used to simplify
future highly accurate computations of highly excited bound states in
two-electron systems.

\begin{center}
   {\bf Acknowledgements}
\end{center}
It is a pleasure to thank David H. Bailey (Berkeley, California) and the
University of Western Ontario for financial support.

\begin{center}
   {\bf Appendix}
\end{center}

The formula obtained by Racah, Rosental and Breit for the field shift is
\begin{eqnarray}
 E^{fs}_M = \frac{4 \pi a^2_0}{Q} \cdot \delta({\bf r}_{eN}) \cdot
 \frac{\gamma + 1}{\Gamma(2 \gamma + 1)} \cdot B(\gamma) \cdot
 \Bigl( \frac{2 Q R}{a_0} \Bigr)^{2 \gamma} \cdot \frac{\delta R}{R}
 \label{Ap1}
\end{eqnarray}
where $Q$ is the nuclear charge, $R$ is the nuclear radius and $\gamma =
\sqrt{1 - \alpha^2 Q^2}$. For the helium atoms we have $\gamma \approx$
0.9998935. Therefore, to a very good accuracy one can assume that $\gamma
\approx 1$. The factor $B(\gamma)$ in Eq.(\ref{Ap1}) is
\begin{eqnarray}
 B(\gamma) = \frac{3}{(2 \gamma + 1) (2 \gamma + 3)} \approx \frac15
\end{eqnarray}
It corresponds to the uniform distribution of the proton density over the
volume of the nucleus. The ratio $\frac{\delta R}{R}$ in the formula
Eq.(\ref{Ap1}) equals unity, if the field shift is determined in respect to
the atom with a point nucleus. Now, in atomic units the formula
Eq.(\ref{Ap1}) takes the form
\begin{eqnarray}
 E^{fs}_M = \frac{8 \pi Q}{5} \cdot \delta({\bf r}_{eN}) \cdot
 \Bigl( \frac{R}{r_e} \frac{r_e}{a_0} \Bigr)^2 \label{Ap2}
\end{eqnarray}
where $r_e = \frac{e^2}{m_e c^2} = \alpha^2 a_0$ is the classical electron
radius. Finally, we obtain the following formula for $E^{fs}_M$
\begin{eqnarray}
 E^{fs}_M = \frac{8 \pi Q}{5} \alpha^4 \cdot \delta({\bf r}_{eN}) \cdot
 \Bigl( \frac{R}{r_e} \Bigr)^2 \label{Ap3}
\end{eqnarray}
This formula can slightly be modified to include other possible
distributions of the proton density in the nuclei. The formula given in the
main text contains an additional factor $f(\lambda) = \frac{1 + \frac13
\lambda}{1 - \frac15 \lambda}$. The formula Eq.(\ref{Ap3}) with such a
factor (i.e. the formula Eq.(\ref{e23A}) from the main text) also provides
the correct answer in those cases when the uniform distribution of the
proton density over the surface of the nucleus is assumed (in this case
$\lambda = 1$). Furthermore, variations of the parameter $\lambda$ in
Eq.(\ref{e23A}) ($0 \le \lambda \le 1$) allow one to describe other possible
distributions of the proton density in light atomic nuclei with $Q \le 6$.

\newpage
%
%
  \begin{table}
   \caption{The total non-relativistic energies $E$ and
            $\langle r_{21} \rangle, \langle \delta_{31} \rangle$,
            $\nu_{31}$ expectation values computed for the excited
            $2^1S(L = 0)-$states of the ${}^{\infty}$He atom (in
            atomic units).}
     \begin{center}
     \begin{tabular}{lllll}
      \hline\hline
$N$ & $E$(${}^{\infty}$He) & $\langle r_{21} \rangle$ & $\langle \delta_{31}
 \rangle$ & $\nu_{31}$ \\
     \hline
 1500 & -2.145 974 046 054 416 890 40 & 5.26969620234182140 &
         1.3094607823 & -2.0000000848 \\

 2000 & -2.145 974 046 054 417 174 01 & 5.26969620234181075 &
         1.3094607799 & -1.9999999787 \\

 2500 & -2.145 974 046 054 417 298 64 & 5.26969620234180607 &
         1.3094607811 & -2.0000000487 \\

 3000 & -2.145 974 046 054 417 342 16 & 5.26969620234180443 &
         1.3094607802  & -1.9999999798 \\

 3500 & -2.145 974 046 054 417 372 44 & 5.26969620234180331 &
         1.3094607801 & -1.9999999784 \\

 4000 & -2.145 974 046 054 417 385 25 & 5.26969620234180281 &
         1.3094607802 & -1.9999999977 \\

 4200 & -2.145 974 046 054 417 391 41 & 5.26969620234180258 &
         1.3094607804 & -2.0000000089 \\
       \hline
 $\infty^{(a)}$ & -2.145 974 046 054 417 415(10) & & & -2.0 \\
       \hline
   & -2.145 974 046 054 419(6)$^{(b)}$ \cite{Drak} & & & \\

   & -2.145 974 04(1)$^{(b)}$ \cite{Acad} & & & \\
  \hline\hline
  \end{tabular}
  \end{center}
${}^{(a)}$The asymptotic value of the total energy (in $a.u.$) \\
${}^{(b)}$The best variational results known from earlier calculations
\cite{Drak} and \cite{Acad} in atomic units.
  \end{table}
  \begin{table}[tbp]
   \caption{The bound state properties $X$ computed for the excited
            $2^1S(L = 0)-$state in the ${}^{\infty}$He, ${}^{4}$He
            and ${}^{3}$He atoms (in atomic units).}
     \begin{center}
     \scalebox{0.63}{%
     \begin{tabular}{cccc}
      \hline\hline
$X$ & ${}^{\infty}$He & ${}^{4}$He & ${}^{3}$He \\
        \hline
$E$ &  -2.14597404605441739141 &
       -2.14567858758314906950 &
       -2.14558192369821233185 \\
$\langle T \rangle$ &
2.14597404605441739081 &
2.14567858758314906890 &
2.14558192369821233125 \\
$\langle V \rangle$ &
-4.29194809320883477957 &
-4.29135717516629813451 &
-4.29116384739642465859 \\
$\eta$ & 4.571$\cdot 10^{-19}$ & 7.8674$\cdot 10^{-19}$ &
9.1373$\cdot 10^{-19}$ \\
     \hline
$\langle r_{31}^{-2} \rangle$ &
0.1437248133044633 &
0.1436882345740985 &
0.1436762703243981 \\
$\langle r_{21}^{-2} \rangle$ &
4.1469390197898792 &
4.1458017822058179 &
4.1454297481103379 \\
$\langle r_{31}^{-1} \rangle$ &
1.13540768612560041 &
1.13525123516046493 &
1.13520005025470815 \\
$\langle r_{21}^{-1} \rangle$ &
0.24968265239356710 &
0.24964776547556185 &
0.24963635362241032 \\
       \hline
$\langle \frac{1}{r_{21} r_{31} r_{32}} \rangle$ &
0.39836585186133074 &
0.39819906128376390 &
0.39814450960354474 \\
$\langle \frac{1}{r_{21} r_{31}} \rangle$ &
0.34063384586003005 &
0.34053958784463326 &
0.34050875587573647 \\
$\langle \frac{1}{r_{31} r_{32}} \rangle$ &
0.56186146745960207 &
0.56170295659588612 &
0.56165110690333470 \\
          \hline
$\frac12 \Bigl[ \langle \frac{r^2_{32}}{r^{3}_{31}} \rangle - \langle
                        \frac{r^2_{21}}{r^{3}_{31}} \rangle \Bigr]$ &
-0.625372574195182710 &
-0.625157759769931583 &
-0.625087487579971712 \\
$\langle \frac{{\bf r}_{31} \cdot {\bf r}_{32}}{r^{3}_{31}} \rangle$ &
-0.0576687311323827964 &
-0.0575321421896994047 &
-0.0574874624526176361 \\
$\langle \frac{{\bf r}_{31} \cdot {\bf r}_{21}}{r^{3}_{31}} \rangle$ &
1.19307641725798262375 &
1.19278337735016376067 &
1.19268751270732578757 \\
    \hline
$\langle r_{31} \rangle$ &
2.973061134389489243 &
2.973491187710339638 &
2.973631893971735653 \\
$\langle r_{21} \rangle$ &
5.269696202341802818 &
5.270450917235814928 &
5.270697843066571909 \\
$\langle r^2_{31} \rangle$ &
16.08923324404050295 &
16.09391391074115758 &
16.09544547281748570 \\
$\langle r^2_{21} \rangle$ &
32.30238037187441050 &
32.31159744722066081 &
32.31461335707651003 \\
       \hline
$\langle r^3_{31} \rangle$ &
108.06042106146845 &
108.10756711191265 &
108.12299522961221 \\
$\langle r^3_{21} \rangle$ &
224.03043954850232 &
224.12604278314834 &
224.15732795934735 \\
$\langle r^4_{31} \rangle$ &
825.75317813369571 &
826.23293858862828 &
826.38995080810075 \\
$\langle r^4_{21} \rangle$ &
1737.415292784614 &
1738.402036368951 &
1738.724968194081 \\
     \hline
$\tau_{31}$ &
0.557144578327034669 &
0.557137512037319818 &
0.557135200602530583 \\
$\tau_{21}$ &
-0.014657043357177236 &
-0.014643315290865667 &
-0.014638824110935725 \\
$\langle f \rangle$ &
2.49080283242230255$\cdot 10^{-2}$ &
2.49079271959434924$\cdot 10^{-2}$ &
2.49078942735313599$\cdot 10^{-2}$ \\
     \hline
$\langle -\frac12 \nabla^2_1 \rangle$ &
1.0729870230272085694 &
1.0726915861654766111 &
1.0725949387331539546 \\
$\langle -\frac12 \nabla^2_3 \rangle$ &
2.1554779104737081945 &
2.1548473559242290248 &
2.1546410836274219596 \\
$\langle {\bf p}_1 \cdot {\bf p}_3 \rangle$ &
-2.1554779104737081945 &
-2.1548473559242290248 &
-2.1546410836274219596 \\
$\langle {\bf p}_1 \cdot {\bf p}_2 \rangle$ &
9.5038644192910556894$\cdot 10^{-3}$ &
9.4641835932758026973$\cdot 10^{-3}$ &
9.4512061611140503550$\cdot 10^{-3}$ \\
$\langle {\bf r}_{21} \cdot {\bf r}_{31} \rangle$ &
16.1511901859372052475 &
16.1557987236103304049 &
16.1573066785382550123 \\
$\langle {\bf r}_{31} \cdot {\bf r}_{32} \rangle$ &
-6.195694189670229434$\cdot 10^{-2}$ &
-6.188481286917282672$\cdot 10^{-2}$ &
-6.186120572076931285$\cdot 10^{-2}$ \\
     \hline
$\langle \delta({\bf r}_{31}) \rangle$ &
1.3094607802 &
1.3089223177 &
1.3087461812 \\
$\nu_{31}$ &
-1.999999997659 &
-1.999725848502 &
-1.999636146310 \\
$\nu^a_{31}$ & -2.0 &
-1.999725850875267686059 &
-1.999636157582479619549 \\
$\langle \delta({\bf r}_{21}) \rangle$ &
8.648433654$\cdot 10^{-3}$ &
8.645119901$\cdot 10^{-3}$ &
8.644035949$\cdot 10^{-3}$ \\
$\nu_{21}$ &
0.4999998171 &
0.4999994703 &
5.0000004784 \\
$\nu^a_{21}$ & 0.5 & 0.5 & 0.5 \\
$\langle \delta_{321} \rangle$ &
0.1755555002 &
0.1753991714 &
0.1753446276 \\
    \hline
$\langle r_{21}^{-3} \rangle_R$ &
-0.04073283492915455 &
-0.04069706227378580 &
-0.04068536285394055 \\
$\langle r_{31}^{-3} \rangle_R$ &
-39.23799206891561 &
-39.21960245645624 &
-39.21358724048685 \\
$\langle r_{21}^{-3} \rangle$ &
0.06794658607018129 &
0.06794071686462781 &
0.06793879648060461 \\
$\langle r_{31}^{-3} \rangle$ &
-22.7828226037360592 &
-22.7711995110946440 &
-22.6896828517925066 \\
  \hline\hline
  \end{tabular}}
  \end{center}
${}^{(a)}$The exact value.
  \end{table}
  \begin{table}[tbp]
   \caption{The total non-relativistic energies $E$ (preliminary results)
            of some excited $n^1S(L = 0)-$states ($n$ = 3, 4) of the
            ${}^{\infty}$He atom (in atomic units).}
     \begin{center}
     \begin{tabular}{|l|l|l|}
     \hline\hline
$N$ & $E$($3^1S-$state) & $E$($4^1S-$state) \\
     \hline
 1500 & -2.061 271 989 740 906 234 & -2.033 586 717 030 718 055 \\

 2000 & -2.061 271 989 740 907 482 & -2.033 586 717 030 722 441 \\

 2500 & -2.061 271 989 740 907 999 & -2.033 586 717 030 723 921 \\

 3000 & -2.061 271 989 740 908 266 & -2.033 586 717 030 724 515 \\

 3500 & -2.061 271 989 740 908 387 & -2.033 586 717 030 724 857 \\

 3800 & -2.061 271 989 740 908 430 & -2.033 586 717 030 725 074 \\
       \hline
 $\infty^{(a)}$ & -2.061 271 989 740 908 48(5) &
 -2.033 586 717 030 725 20(7) \\
       \hline
 \cite{Drak}$^{(b)}$ & -2.061 271 989 740 911(5) &
     -2.033 586 717 030 72(1) \\
   \hline\hline
  \end{tabular}
  \end{center}
${}^{(a)}$The asymptotic value of the total energy (in $a.u.$) \\
${}^{(b)}$The best variational results known from earlier calculations
\cite{Drak} in atomic units.
  \end{table}
\end{document}